\documentclass[%
 aip,
 jmp,%
 amsmath,amssymb,
 reprint,%
]{revtex4-2}

\bibpunct{[}{]}{;}{n}{}{}

\usepackage{graphicx}
\usepackage{dcolumn}
\usepackage{bm}

\usepackage{subcaption}
\usepackage{physics}

\begin{document}

\preprint{APS}

\title{General Relativistic Mean-field Dynamo Model for Proto-neutron Stars}

\author{K. Franceschetti}
\affiliation{Dipartimento di Fisica e Astronomia, Universit\`a degli Studi di Firenze, Italy}

\author{L. \surname{Del Zanna}}\email{luca.delzanna@unifi.it}
\affiliation{Dipartimento di Fisica e Astronomia, Universit\`a degli Studi di Firenze, Italy}
\affiliation{INAF - Osservatorio Astrofisico di Arcetri, Firenze, Italy}
\affiliation{INFN - Sezione di Firenze, Italy}

\date{\today}

\begin{abstract}
Neutron stars, and magnetars in particular, are known to host the strongest magnetic fields in the Universe. The origin of these strong fields is a matter of controversy. In this preliminary work, via numerical simulations, we study, for the first time in non-ideal general relativistic magnetohydrodynamic (GRMHD) regime, the growth of the magnetic field due to the action of the mean-field dynamo due to sub-scale, unresolved turbulence. The dynamo process, combined with the differential rotation of the (proto-)star, is able to produce an exponential growth of any initial magnetic seed field up to the values required to explain the observations. By varying the dynamo coefficient we obtain different growth rates. We find a quasi-linear dependence of the growth rates on the intensity of the dynamo. Furthermore, the time interval in which exponential growth occurs and the growth rates also seems to depend on the initial configuration of the magnetic field.
\end{abstract}


\maketitle

\section{Introduction}\label{sec1}
A neutron star (NS) is the smallest and densest star in the Universe, with radius of the order of 10 kilometers and mass between about 1.1 and 2.17 solar masses (e.g.,
\cite{Ozel(2012),Rezzolla(2018),Cromartie(2019)}). 
Together with black holes, it is one of the possible compact objects in which the core of a giant star - a star with a mass greater than eight solar masses - can collapse at the end of its life cycle.

NSs show a vast and diverse phenomenology. Their magnetic fields are between $10^8$ and $10^{15}$ times stronger than Earth's one. Intense magnetic fields can deform the star, making it more or less oblate \cite{Kiuchi(2008),Ciolfi(2009),Frieben(2012),Pili(2014),Bucciantini(2015)}), and give observable signals of gravitational waves (GWs), especially during merging events like GW170817, the first gravitational wave signal due to the merging of two neutron stars observed by the LIGO and Virgo detectors on 17 August 2017 \cite{Rezzolla(2018),DallOsso(2009),Abbott(2017)}.

A newly born, fast rotating and ultra-magnetized proto-NS can play as central engine driving the energetic emission of both Long and Short Gamma-Ray Bursts (GRBs) \cite{DallOsso(2009),Uso(1992),Bucciantini(2009),Bucciantini(2012),Moesta(2015),Ciolfi(2019)}, the luminous electromagnetic events known to occur in the Universe. Once formed, rapidly rotating and strongly magnetized NSs would lose rotational kinetic energy in a short time, on a timescale of seconds or less. A relativistic plasma, driven the combination of the rotation with the magnetic field, flows away from the star, and its emission of X-ray and $\gamma$-ray in the photosphere can reproduce the observational characteristics of a GRB \cite{Uso(1992)}.

The origin of these strong magnetic fields of neutron stars - and in particular those of magnetars, which have magnetic fields of the order of $10^{14} - 10^{15}$ gauss (G) - is a matter of controversy and many evolutionary scenarios have been proposed so far. In the original magnetar model \cite{Duncan(1992)} the strength of the magnetic field is the result of the growth of the seed field amplified by the dynamo mechanism - a process through which a rotating, convecting, and electrically conducting fluid can generate a magnetic field by self-inductive action converting kinetic energy into magnetic energy - in the proto-neutron star (PNS), the remnant of the supernova explosion which represents the first phase of life of a NS. The amplification of the magnetic field by a turbulent dynamo can occur either by a convective dynamo (e.g., \cite{Thompson(1993),Raynaud(2020)}) or the magnetorotational instability (MRI, e.g., \cite{Akiyama(2003),Obergaulinger(2009),Reboul-Salze(2020)}).

Most of the studies done so far have focused on the so-called mean-field dynamo theory (e.g., \cite{Bonanno(2003),Brandenburg(2005),Naso(2008),Raynaud(2020)}), in which the magnetic field is amplified by an electromotive force generated by the coupling between velocity and magnetic field small scale turbulent fluctuations (which we do not solve and must be 3D to generate dynamo) - almost certainly induced by MRI. However, all the studies conducted so far have been done using classical dynamo equations. A fully covariant generalization of equations in the general relativistic magnetohydrodynamical (GRMHD) regime was first proposed by Bucciantini \& Del Zanna in \cite{Bucciantini(2013)} using the so-called 3+1 formalism.

In this preliminary work we examine for the first time with numerical simulations in non-ideal GRMHD the amplification of a seed field by the mean-field dynamo in the so-called kinematic approximation, in which the star is assumed to be in stationary hydrostatic equilibrium. As a result, the velocity field is taken as pre-assigned and time-independent. Our goal is to verify the feasibility of the proposed mechanism and to estimate the average growth rate of the magnetic field, investigating its dependence on the intensity of the dynamo.

This paper is organized as follows. In Section \ref{sec2}, we briefly present the clasical dynamo theory and its fully covariant generalization proposed by Bucciantini \& Del Zanna. In Section \ref{sec3}, we describe our model and its initialization. In Section \ref{sec4} we show and discuss the results of our numerical simulations. Finally, we conclude in Section \ref{sec5}.

\section{The Mean-field Dynamo}\label{sec2}
The dynamo theory describes the process through which a rotating, convecting, and electrically conducting fluid can generate a magnetic field by self-inductive action converting kinetic energy into magnetic energy and maintain it over astronomical time scales. In particular, motion through a magnetic field induces an electric field that generates a current according to Ohm's law. The result is a highly nonlinear system, in which this current in turn produces a magnetic field which on the one hand generates a new electric field through Faraday's law and on the other can oppose motion by means of the Lorentz force. However, the onset of dynamo action can be studied in the linear approximation, i.e., the velocity field is assumed to be given (kinematic problem).

We will concentrate on the process known as mean-field dynamo in the magnetohydrodynamical (MHD) regime, which has been applied to a large number of astrophysical contexts, as the Sun \cite{Parker(1955)}, neutron stars (e.g., \cite{Bonanno(2003)}), and accretion disks (e.g., \cite{Bugli(2014),Tomei(2020)}).

In the following we assume a signature $\qty(-, +, +, +)$ for the space-time metric and we use Greek letters (running from 0 to 3) for 4D space-time components, and Latin letters (running from 1 to 3) for 3D spatial components. Moreover, we set $G = c = M_{\odot} = 1$ (where $M_{\odot}$ is the solar mass). Finally, for the electromagnetic quantities we make use of the Lorentz-Heaviside units, where $\varepsilon_0 = \mu_0 = 1$.

\subsection{Classical Theory}
In the classical MHD, the time evolution of the electric field can be neglected in the Ampère-Maxwell equation, which can thus replaced by the Ampère's law $\curl\vb{B} = \vb{J}$. Substituting Ohm's law
\begin{equation}
\vb{J} = \sigma\vb{E}' \qquad \vb{E}' = \vb{E}+\vb{U}\cp\vb{B}
\end{equation}
- where $\sigma$ is the electric conductivity, $\vb{U}$ is the fluid velocity and $\vb{E}'$ is the electric field in the frame comoving with the fluid - into the Faraday's law of induction, one can write a single evolution equation for $\vb{B}$, which is called the induction equation:
\begin{equation}
\partial_t \vb{B} = \curl\qty(\vb{U}\cp\vb{B}-\eta_d \vb{J})
\end{equation}
where $\partial_t = \pdv*{t}$ and $\eta_d = \sigma^{-1}$ is the magnetic diffusivity. As mentioned in section \ref{sec1}, in the mean-field dynamo theory, if small scale turbulent fluctuations of the magnetic field are well developed, the coupling between velocity and magnetic field fluctuations generates an electromotive force that can be written as
\begin{equation}
\langle{\var\vb{v}\cp\var\vb{B}}\rangle  = \alpha_{turb} \vb{B} - \beta_{turb} \vb{J}
\label{eq3}
\end{equation}

The first term is the so-called $\alpha$-term, describing the creation of poloidal fields starting from toroidal fields, while the second term describes the strongly enhanced dissipation of the field by the turbulence. In general both $\alpha_{turb}$ and $\beta_{turb}$ will be tensors, however we will focus here on the isotropic case where they can be dealt with as scalars. Thanks to the combined action of these two terms, this electromotive force amplifies the initial magnetic field according to the mechanism known as $\alpha\Omega$-dynamo, where the $\Omega$ effect refers to the amplification of the toroidal field by shear (i.e., differential rotation). It is possible to prove (e.g., \cite{Bucciantini(2013)}) that the comoving electric field can be rewritten as
\begin{equation}
\vb{E}' = \xi\vb{B} + \eta\vb{J}
\end{equation}
where $\xi = -\alpha_{turb}$ and $\eta = \eta_d + \beta_{turb}$.

\subsection{Mean-field Dynamo in 3+1 Resistive GRMHD}\label{sec2.2}
As mentioned in section \ref{sec1}, the first fully covariant generalization of Equation \eqref{eq3} in GRMHD was first proposed by Bucciantini \& Del Zanna in \cite{Bucciantini(2013)} (see also \cite{DelZanna(2018)}). Extending the usual form adopted for resistive GRMHD, the expression for the comoving electric field appears to be
\begin{equation}
e^{\mu} = \xi b^{\mu} + \eta j^{\mu}
\label{eq5}
\end{equation}
where $j^{\mu}$ and $b^{\mu}$ are the comoving conduction current and the comoving magnetic field, respectively. While the covariant form of equations is very important - and quite elegant - from a theoretical point of view, it does not allow one to think clearly about the evolution in time of a physical system and to give a proper physical meaning to the various relativistic quantities. Moreover the covariant formalism is difficult to use in numerical analysis. Therefore, instead of using covariant fields, time and space are divided by numerical integration. Using the 3+1 formalism - in which the line element can be written as
\begin{equation}
\dd s^2 = -\alpha^2 \, \dd t^2 + \gamma_{ij}\qty(\dd x^i + \beta^i \dd t)\qty(\dd x^j + \beta^j \dd t)
\end{equation}
where $\alpha$ is known as the lapse function, $\vb*{\beta}$ as the shift vector, and $\gamma_{ij}$ is the 3-metric, used to raise/lower the indexes of any spatial three-dimensional vector or tensor - Maxwell's equations become
\begin{equation}
\begin{aligned}
& \partial_t \qty(\sqrt{\gamma} \, E^i) - \sqrt{\gamma} \epsilon^{ijk}\partial_j \qty(\alpha B_k - \epsilon_{klm} \beta^l E^m) = - \sqrt{\gamma} \, \qty(\alpha J^i - q\beta^i) \\
& \partial_t \qty(\sqrt{\gamma} \, B^i) + \sqrt{\gamma} \epsilon^{ijk}\partial_j \qty(\alpha E_k + \epsilon_{klm}\beta^l B^m) = 0 \\
& \partial_i \qty(\sqrt{\gamma} E^i) = \sqrt{\gamma} \, q \\
& \partial_i \qty(\sqrt{\gamma} B^i) = 0
\end{aligned}
\end{equation}
where $\gamma = \det\gamma_{ij}$ and $\epsilon_{ijk}$ are respectively the determinant and the Levi-Civita pseudo-tensor of the 3-metric. We note that we must integrate the electric field over time and the current $\vb{J}$ appears as a source term. Writing Equation \eqref{eq5} in the 3+1 form we find 
\begin{equation}
J^i = qv^i + \sigma_E \, \Gamma \, \qty(E^i + \epsilon^{ijk}v_j B_k - \qty(v_k E^k)v^i) + \sigma_B \, \Gamma \, \qty(B^i - \epsilon^{ijk}v_j E_k - \qty(v_k B^k)v^i)
\end{equation}
where $\Gamma$ is the Lorentz factor, $q$ is the electric charge density, $\sigma_E = 1/\eta$, and $\sigma_B = -\xi/\eta$. Due to the presence of $\eta$ in the denominator, stiff source terms arise in the equation for the evolution of the electric field, requiring some sort of implicit treatment (e.g., \cite{DelZanna(2016)}).

\section{Model and Numerical Set-up}\label{sec3}
The fluid configuration of our PNS is in axisymmetric equilibrium and has been modeled via the \texttt{XNS} code \cite{Bucciantini(2011),Pili(2014),Pili(2017)} with an uniform grid in both radial (256 points in the range $\qty[0,30]$ (geometrized units), with the star resolved in about half of the points) and $\theta$ (64 points in the range $\qty[0,\pi]$) directions. A polytropic relationship between pressure and density is, i.e.,
\begin{equation}
p = K\rho^{\gamma}
\end{equation}
where $p$ is the thermal pressure, $\rho$ is the mass density, and $\gamma$ is the adiabatic index. Table \ref{table1} shows
all parameters defining the fluid configurations. Here $\rho_c \approx 7.9\times 10^{14}$ g/cm$^3$ is the central density, $\Omega_c \approx 5.3$ kHz is the central rotation rate and $\Omega_{eq} \approx 2.0$ kHz is the equatorial one, $R_{eq} \approx 12.4$ km is the equatorial radius of the star, and $A$ is a measure of the differential rotation rate and is such that (see \cite{Bucciantini(2011)})
\begin{equation}
A^2 \, \qty(\Omega_c - \Omega) = \dfrac{\mathcal{R}^2 \, \qty(\Omega + \beta^{\phi})}{\alpha^2 - \mathcal{R}^2 \, \qty(\Omega + \beta^{\phi})^2}
\end{equation}
where $\mathcal{R} = \sqrt{\gamma_{\phi\phi}}$. The corresponding central spin period is $P_c \approx 1.2$ ms $\ll P_{crit}$, where $P_{crit} \sim 1-1.5$ s is the the critical value corresponding to the upper limit of the rotation period of the PNS which allows the turbulent mean-field dynamo to develop \cite{Bonanno(2003)}. The choice of this high central rotation rate is due to the desire to maximize the $\Omega$ effect. Note that the equation of state (EoS) and the rotation profile we use are more appropriate for an NS than for a PNS (e.g., \cite{Reboul-Salze(2020)}). However, the aim of this preliminary work is to verify whether the dynamo model could be a realistic scheme for generating proto-magnetar, taking into account, for the first time, non-ideal effects in general relativity. The results can be easily extended to PNS, for which one should change the EoS and rotation profile.

\begin{table}[h!]
\caption{Parameters of the fluid model in geometrized units ($c=G=M_{\odot}=1$). The corresponding values in cgs units are reported in the text.}
\label{table1}
\begin{ruledtabular}
\begin{tabular}{ccccccc}
$\boldsymbol{K}$ & $\boldsymbol{\gamma}$ & $\boldsymbol{\rho_c}$ & $\boldsymbol{\Omega_c}$ & $\boldsymbol{A^2}$ & $\boldsymbol{\Omega_{eq}}$ & $\boldsymbol{R_{eq}}$\\
\hline
100 & 2 & $1.28\times 10^{-3}$ & $2.633\times 10^{-2}$ & 70 & $9.660\times 10^{-3}$ & 8.38\\
\end{tabular}
\end{ruledtabular}
\end{table}

The study of the time evolution of the system is performed via the \texttt{ECHO} code \cite{DelZanna(2007)}. Therefore, at the initial time (i.e., at $t=0$), in addition to the configuration of the stellar fluid just described, a small magnetic field (added ``by hand'') and the profiles of the $\xi$ and $\eta$ parameters must be inserted (see below).

For the magnetic field we use a purely toroidal initial configuration. As shown in \cite{Bucciantini(2011)}, the toroidal field can be written as
\begin{equation}
B^{\phi} = \dfrac{K_m \, \qty(\alpha^2 \, \mathcal{R}^2 \, \rho h)^m}{\alpha\mathcal{R}^2}
\end{equation}
where $h = 1 + \gamma/(\gamma-1)p/\rho$ is the specific enthalpy, $K_m$ is the toroidal magnetization constant, which regulates the strength of the magnetic field (more specifically the magnetic flux through the meridional plane), and $m\geq 1$ is the toroidal magnetization index (related to the distribution of the magnetic field inside the star). We choose $K_m = 5\times 10^{-5}$ and $m = 1.5$ so as to have, in physical units, an initial magnetic field of the order of $10^{12}$ G, i.e., the typical value inferred for pulsars. The electric field was instead initialized using the relation
\begin{equation}
\vb{E} = -\vb{v}\cp\vb{B}
\end{equation}
of the ideal MHD. Therefore, the initial electric field is zero with a purely toroidal initial field (as assumed here by us), but not with an initial poloidal field, as we will assume in Section \ref{sec4.2}.

In order to quantify the dynamo action, we can introduce the two characteristic (dimensionless) numbers (e.g., \cite{Brandenburg(2005)})
\begin{equation}
C_{\xi} = \dfrac{\xi_0 R}{\eta} \qquad C_{\Omega} = \dfrac{\Delta\Omega R^2}{\eta}
\end{equation}
where $\Delta\Omega = \Omega_c - \Omega_{eq}$, $R = R_{eq}$, and $\xi_0$ is the maximum value of $\xi$. These numbers may be regarded as magnetic Reynolds numbers based on the intensity of the $\alpha$-effect and the differential rotation, respectively. The $\xi$ and $\eta$ profiles are chosen so that the diffusion and dynamo processes occur only within the star. While we assume constant $\eta$ within the star, the profile of $\xi$ is similar to that used by Bonanno et al. in \cite{Bonanno(2003)}, namely
\begin{equation}
\xi =
\begin{cases}
\xi_0 \, f_{\xi}\qty(r,\theta) \, \mathcal{Q}_{\xi}\qty(\vb{B}) & \text{ inside the star} \\[1pc]
0 & \text{ in the atmosphere}
\end{cases}
\label{eq14}
\end{equation}
where
\begin{equation}
f_{\xi}\qty(r,\theta) = \dfrac{\mathrm{s}}{2}\sin\theta \, \qty[1+\erf\qty(\dfrac{r-R_c}{\Delta R})]
\end{equation}
with erf the error function,
\begin{equation}
\mathrm{s} = \mathrm{s}\qty(\theta) =
\begin{cases}
+1 & \mbox{ for } \dfrac{\pi}{2}-\theta \geq 0 \\[1pc]
-1 & \mbox{ for } \dfrac{\pi}{2}-\theta < 0
\end{cases}
\end{equation}
and $\Delta R = 0.025 R$. The function
\begin{equation}
\mathcal{Q}_{\xi}\qty(\vb{B}) = \qty[1+\dfrac{\vb{B}^2}{B_{eq}^2}]^{-1}
\end{equation}
is known as quenching function. In the kinematic approximation, quenching is necessary to prevent the infinite growth of the magnetic field due to the absence of the fluid feedback. Here $B_{eq}$ is the equipartition magnetic field with respect to the kinetic energy density of turbulent motions. We assumed
\begin{equation}
B_{\text{eq}}^2 = \alpha_q \, p
\end{equation}
where $\alpha_q$ is a constant and $p$ is the thermal pressure.
As mentioned above, relation \eqref{eq14} is similar to the one assumed by Bonanno et al. in \cite{Bonanno(2003)}. The difference is that we used the sine function with respect to the cosine function to have the maximum dynamo parameter near the equator of the star. The antisymmetry across the equatorial plane is guaranteed by the $\mathrm{s}$ function.

From a physical point of view, relation \eqref{eq14} implies that the star is divided into two regions: an innermost one where the dynamo is negligible and an outermost one where it is maximum. This two regions present two different instabilities: a convective one is active in the inner regions of the star (of radius $R_c$) and a so-called neutron-finger instability, due to the lepton number gradients (see \cite{Bonanno(2003),Naso(2008)}), in the outer regions. The two regions are separated by a thin layer - named tachocline - of thickness $\Delta R$ through which the physical properties of the two regions are supposed to vary in a smooth way. In general, $R_c$ varies over time. Taking $R_c$ constant during the entire evolution is equivalent to assume that $R_c$ varies in much longer times than those in which the magnetic field grows exponentially due to the dynamo.

We choose five different values of $\xi_0$. The list of runs with the corresponding parameters is reported in Table \ref{table2}. The value of $\alpha_q$ is chosen so that, in physical units, $B_{eq} \sim 10^{15}$ G, i.e., the standard value for magnetars. Recall that in this preliminary work we decided to vary only the most important quantity, i.e., the maximum value of $\xi$ (and therefore $C_{\xi}$) for the same star and the same profiles of $\xi$ and $\eta$.

\begin{table}[h!]
\caption{Parameters of the fluid model in geometrized units ($c=G=M_{\odot}=1$).}
\label{table2}
\begin{ruledtabular}
\begin{tabular}{ccccccc}
& $\boldsymbol{\xi_0}$ & $\boldsymbol{\eta}$ & $\boldsymbol{C_{\xi}}$ & $\boldsymbol{C_{\Omega}}$ & $\boldsymbol{R_c/R}$ & $\boldsymbol{\alpha_q}$\\
\hline
Run1 & $1.0\times 10^{-2}$ & $1.0\times 10^{-3}$ & $0.084\times 10^3$ & $1.170\times 10^3$ & 0.4 & $1.0\times 10^{-5}$\\
Run2 & $5.0\times 10^{-2}$ & $1.0\times 10^{-3}$ & $0.419\times 10^3$ & $1.170\times 10^3$ & 0.4 & $1.0\times 10^{-5}$\\
Run3 & $1.0\times 10^{-1}$ & $1.0\times 10^{-3}$ & $0.838\times 10^3$ & $1.170\times 10^3$ & 0.4 & $1.0\times 10^{-5}$\\
Run4 & $2.0\times 10^{-1}$ & $1.0\times 10^{-3}$ & $1.676\times 10^3$ & $1.170\times 10^3$ & 0.4 & $1.0\times 10^{-5}$\\
Run5 & $3.0\times 10^{-1}$ & $1.0\times 10^{-3}$ & $2.514\times 10^3$ & $1.170\times 10^3$ & 0.4 & $1.0\times 10^{-5}$\\
\end{tabular}
\end{ruledtabular}
\end{table}

\section{Numerical Results}\label{sec4}
In this section we show the results of our $\alpha\Omega$-dynamo simulations performed via the \texttt{ECHO} code using a third order scheme in time and a fifth order scheme in space. Following the prescription of Tomei et al. in \cite{Tomei(2020)}, we define the average on the star of any quantity $f = f\qty(r,\theta)$ as
\begin{equation}
\langle{f}\rangle = \dfrac{\iint_{\mathcal{D}} \dd r \, \dd\theta \, \alpha\sqrt{\gamma} f}{\iint_{\mathcal{D}} \dd r \, \dd\theta \, \alpha\sqrt{\gamma}}
\end{equation}
where $\mathcal{D}$ identifies the region inside the star, $\alpha$ and $\gamma = \det\gamma_{ij}$ are the lapse function and the determinant of the 3-metric defined in Section \ref{sec2.2}, respectively. The presence of the factor $\alpha$ is commonly invoked but still debated. For a better understanding, in this section we report the results in physical units.

\begin{figure}[h!]
\centering
\includegraphics[scale=0.5]{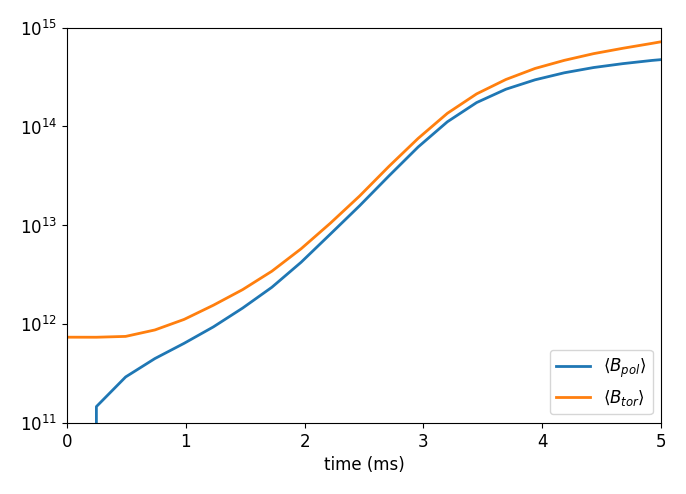}
\caption{Time evolution in the Run1 for the average value of poloidal ($B_{pol}$) and toroidal ($B_{tor}$) components of the magnetic field in the first 5 ms. Values in Gauss (G) along the vertical axis.}
\label{fig1}
\end{figure}

Figure \ref{fig1} shows the time evolution of the average poloidal and toroidal components of the magnetic field in the Run1. We can see that a poloidal field does not arise immediately, but only after a short phase in which the toroidal field remains constant. At this point the dynamo starts and, after a transient phase, supports the exponential amplification of the two components up to $\sim 3$ ms, when the dynamo action is limited by the quenching effect. At the end of the exponential growth phase, the strength of the toroidal component is slightly higher than that of the poloidal one. The time interval in which the dynamo operates is much smaller than that on which the evolution of the neutron-finger instability occurs, which is of the order of seconds \cite{Miralles(2000)}, in support of the hypothesis of constant $R_c$ during the action of the dynamo.

\begin{figure}[h!]
\centering
\begin{subfigure}{0.45\linewidth}
\includegraphics[scale=0.45]{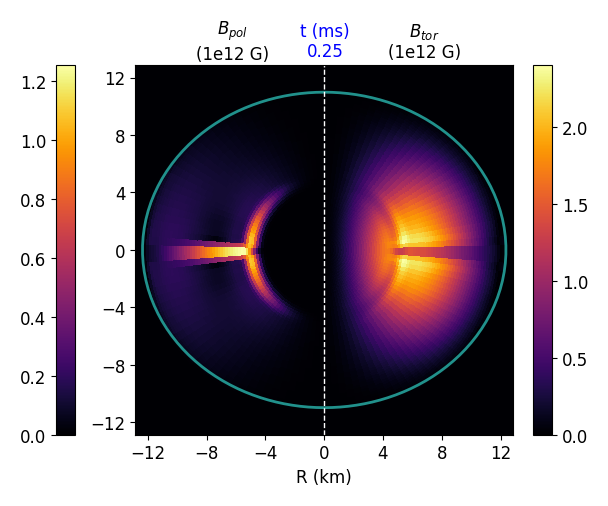}
\caption{\label{fig2a}}
\end{subfigure}
\begin{subfigure}{0.45\linewidth}
\includegraphics[scale=0.45]{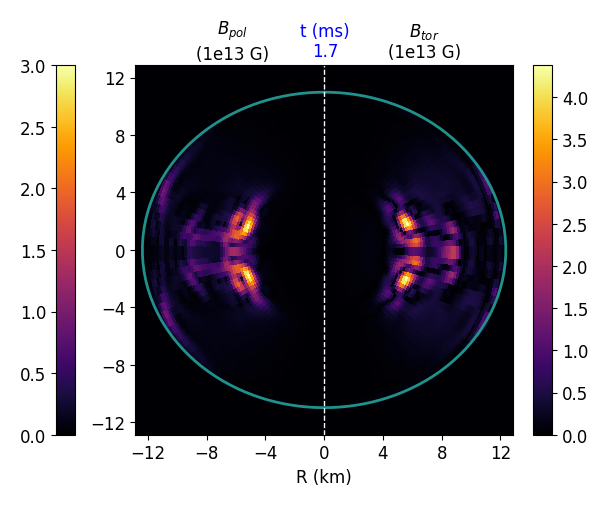}
\caption{\label{fig2b}}
\end{subfigure}
\par\bigskip
\begin{subfigure}{0.45\linewidth}
\includegraphics[scale=0.45]{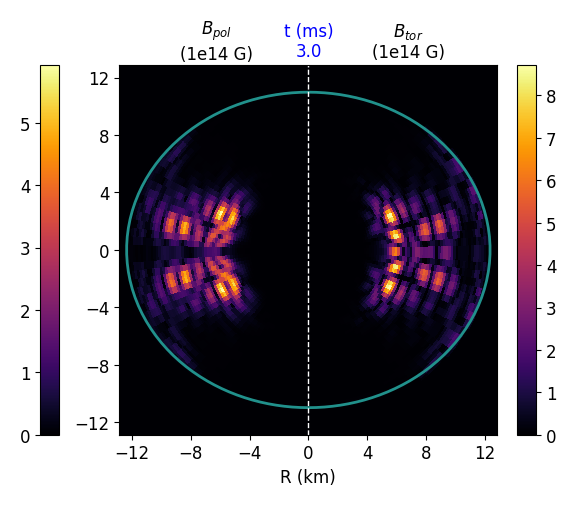}
\caption{\label{fig2c}}
\end{subfigure}
\begin{subfigure}{0.45\linewidth}
\includegraphics[scale=0.45]{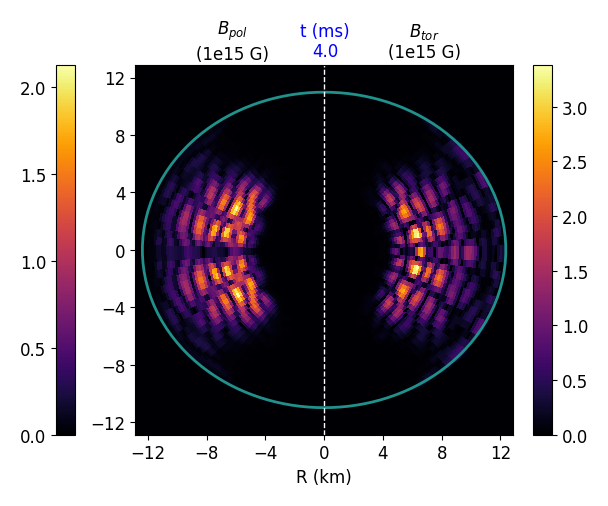}
\caption{\label{fig2d}}
\end{subfigure}
\bigskip
\caption{Color maps of the poloidal ($B_{pol}$, left side of each figure) and toroidal ($B_{tor}$, right side of each figure) components of the magnetic field  at four different times for Run1. The blue solid curve is the surface of the star. In the colorbar, values are in order of (\textbf{a}) $10^{12}$ G; (\textbf{b}) $10^{13}$ G; (\textbf{c}) $10^{14}$ G; (\textbf{d}) $10^{15}$ G.}
\label{fig2}
\end{figure}

The time distribution on the $xz$-plane (any meridional cut of the star) of the components of the magnetic field is shown - at four different times - in Figure \ref{fig2}. We can see that the poloidal field arise near the tachocline and in the outer region of the star (where the dynamo parameter is maximum), mainly near the equator (Figure \ref{fig2a}). During the dynamo action, magnetic islands corresponding to the linear dynamo modes migrate towards the poles and the surface of the star while growing in amplitude (Figures \ref{fig2b}--\ref{fig2c}). This is similar to the migration of sunspots towards the equatorial plane on the surface of the Sun, but in the opposite direction, due to the fact that we have set $\xi = -\alpha_{turb}$ (see Section \ref{sec2}). During the post-exponential growth phase, the field is mainly located - along the $z$-axis - in a region about $0.6$ times the polar diameter (Figure \ref{fig2d}). This evolution is explained as follows. Due to the presence of curls in Maxwell's equations, the gradients of the various quantities have a strong impact on the time evolution of the system. In fact already at small $t$ the magnetic field only grows where there are the gradients of $\xi$, which are stronger than any other in play. Then the various components of the magnetic field are created, amplified and migrate, but the ``imprint'' of the gradients of $\xi$ remains.

\subsection{Dependence on the $\alpha$-dynamo Number}
We now investigate the dependence of the results on the $\alpha$-dynamo number $C_{\xi}$. The list of runs with the corresponding growth rates for both poloidal ($\omega_P$) and toroidal ($\omega_T$) components is reported in Table \ref{table3}.

\begin{table}[h!]
\caption{Exponential growth rates (in ms$^{-1}$) for both poloidal ($\omega_P$) and toroidal ($\omega_T$) components of the magnetic filed for all runs. The corresponding value of the $\alpha$-dynamo number $C_{\xi}$ is also reported.}
\label{table3}
\begin{ruledtabular}
\begin{tabular}{cccc}
& $\boldsymbol{C_{\xi}}$ & $\boldsymbol{\omega_P}$ [\textbf{ms}$\boldsymbol{^{-1}}$] & $\boldsymbol{\omega_T}$ [\textbf{ms}$\boldsymbol{^{-1}}$]\\
\hline
Run1 & $0.084\times 10^3$ & $2.75\pm 0.02$ & $2.67\pm 0.03$\\
Run2 & $0.419\times 10^3$ & $34.6\pm 0.2$ & $33.5\pm 0.1$\\
Run3 & $0.838\times 10^3$ & $45.8\pm 0.3$ & $45.9\pm 0.2$\\
Run4 & $1.676\times 10^3$ & $152\pm 2$ & $146\pm 1$\\
Run5 & $2.514\times 10^3$ & $169\pm 4$ & $163\pm 3$\\
\end{tabular}
\end{ruledtabular}
\end{table}

\begin{figure}[h!]
\centering
\includegraphics[scale=0.5]{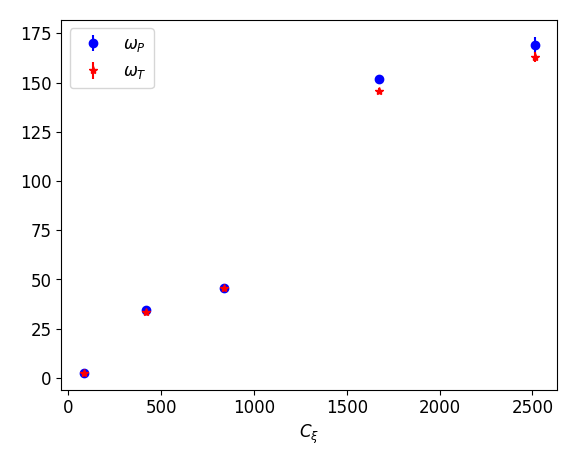}
\caption{Dependence of growth rates $\omega_P$ (blue squares) and $\omega_T$ (red stars) on the dynamo number $C_{\xi}$. Values in ms$^{-1}$ along the vertical axis.}
\label{fig3}
\end{figure}

Figure \ref{fig3} shows the dependence of the exponential growth rates with the dynamo number $C_{\xi}$. Blue (red) squares (stars) for the poloidal (toroidal) field. We observe that both rates follow a quasi-linear dependence on the dynamo number, as expected, with a slope of about $\qty(7\pm 1)\times 10^{-2}$ ms$^{-1}$. For small values of $C_{\xi}$ there is not much difference between the two rates, while for large values of $C_{\xi}$ a slight difference is visible.

As can be seen from the data shown in Table \ref{table3}, the most realistic values for the dynamo parameter are those used for the first runs (or lower). The highest values, used in the latest runs, are used for the sole purpose of determining the trend of growth rates on the dynamo parameter $C_{\xi}$ and a possible drastic variation in the slope (or even a non-linear dependence) for these limit values. In the near future we plan to repeat the study for lower values of $C_{\xi}$, in order to confirm or not this quasi-linear trend of growth rates even for less intense dynamo and the weight that the higher values of $C_{\xi}$ had on the estimate of the slope found in this work.

\subsection{Dependence on the Initial Configuration of the Magnetic Field}\label{sec4.2}
\begin{figure}[h!]
\centering
\includegraphics[scale=0.5]{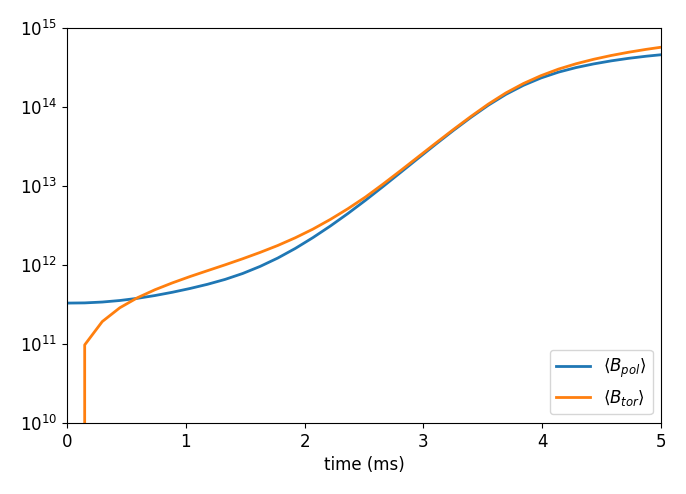}
\caption{Time evolution in the Run1 for the average value of poloidal ($B_{pol}$) and toroidal ($B_{tor}$) components of the magnetic field in the first 5 ms in the case of a purely dipolar initial magnetic field. Values in Gauss (G) along the vertical axis.}
\label{fig4}
\end{figure}

We repeat here Run1 with an initial configuration for the magnetic field different from that used for the previous runs in order to show that the dynamo occurs independently of the initial configuration of the magnetic field and to verify whether or not there is a dependence of the growth rates on the latter. To do this, we consider a purely dipolar instead than toroidal magnetic field. The field is initialized using the analytical solution found by Mastrano et al. in the Newtonian
case in \cite{Mastrano(2013),Mastrano(2015)}, namely
\begin{equation}
\vb{\tilde{B}}_{dip} =
\begin{cases}
\dfrac{35}{4}\qty[\qty(1-\dfrac{6}{5}d^2+\dfrac{3}{7}d^4)\cos\theta \, \vu{e}_r - \qty(1-\dfrac{12}{5}d^2+\dfrac{9}{7}d^4)\sin\theta \, \vu{e}_{\theta}] & \text{for } d < 1 \\[2pc]
\dfrac{2\cos\theta \, \vu{e}_r + \sin\theta \, \vu{e}_{\theta}}{d^3} & \text{for } d \geq 1
\end{cases}
\end{equation}
where $d = r/R_p$ - with $R_p$ the polar radius of the star -, and $\vb{\tilde{B}} = \vb{B}/B_0$, where $B_0$ sets the strength of the field overall, is continuous at $d = 1$. Our definition of $d$ is due to the fact that in our case the star is not completely spherical, but is slightly oblate, as can be seen in Figure \ref{fig2}, and we decided to study the evolution of the system exclusively within the star, where the quantities of interest are significantly appreciable. We choose $B_0 = 10^{-9}$ (in geometrized units), corresponding to a maximum value of the magnetic field of the order of $10^{12}$ G. Figure \ref{fig4} shows the time evolution of the average poloidal and toroidal components of the magnetic field. We see that the exponential growth phase begins and ends with a small delay compared to the previous case. Furthermore, just like in the previous case, in the post-exponential growth phase the toroial field has a higher intensity than the poloidal one.

The exponential growth rates are reported in Table \ref{table4}, where those for the purely toroidal initial configuration of the magnetic field are also reported. We can see that the growth rates in the case of purely dipolar initial field are lower than those obtained with the purely toroidal initial configuration.

\begin{table}[h!]
\caption{Exponential growth rates (in ms$^{-1}$) for both poloidal ($\omega_P$) and toroidal ($\omega_T$) components of the magnetic filed for Run1. The subscript $_t$ ($_d$) is for the purely toroidal (dipolar) initial configuration of the magnetic field.}
\label{table4}
\begin{ruledtabular}
\begin{tabular}{cccc}
& $\boldsymbol{C_{\xi}}$ & $\boldsymbol{\omega_P}$ [\textbf{ms}$\boldsymbol{^{-1}}$] & $\boldsymbol{\omega_T}$ [\textbf{ms}$\boldsymbol{^{-1}}$]\\
\hline
Run1$_t$ & $0.084\times 10^3$ & $2.75\pm 0.02$ & $2.67\pm 0.03$\\
Run1$_d$ & $0.084\times 10^3$ & $2.69\pm 0.02$ & $2.64\pm 0.01$\\
\end{tabular}
\end{ruledtabular}
\end{table}

\section{Summary and Conclusions}\label{sec5}
In this work we investigate, for the first time by means of non-ideal axisymmetric GRMHD simulations, the mean-field dynamo process operating in proto-neutron stars. Our simulations are performed in the so-called kinematic approximation, in which the star is assumed to be in hydrostatic equilibrium at all times. Our star is assumed to have an internal structure described by a polytropic equation of state and to be rapidly rotating - with a central spin period of $P_c \sim P_{crit}/1000$, where $P_{crit}$ is the upper limit of the rotation period of the star which allows the dynamo to develop (see \cite{Bonanno(2003)}) - to maximize the $\Omega$-effect. We also assume a purely toroidal initial configuration for the magnetic field with a strength of the order of $10^{12}$ G.

Since a proto-neutron star is subject to two substantially different instabilities - a convective one active in the inner regions of the star, where the dynamo is probably negligible, and a neutron-finger instability in the outer regions, where the mean-field dynamo can operate \cite{Thompson(1993),Bonanno(2003)} - we assume, for the dynamo parameter, the profile defined by Equation \eqref{eq14}, similar to that assumed by Bonanno et al. in \cite{Bonanno(2003)}. With this model we assume that the relevant physical properties of the two regions vary across a thin layer named tachocline. We choose five different values for the maximum value of the dynamo parameter, while the turbulent magnetic diffusivity is assumed constant within the star.

The study of the evolution of electric and magnetic fields under the action of the dynamo is performed in a stationary spacetime in spherical coordinates via the \texttt{ECHO} code \cite{DelZanna(2007)} in the upgraded version to include non-ideal resistive and dynamo effects in the Ohm's law \cite{Bucciantini(2013),DelZanna(2018)}. We find that the exponential growth rates follow a quasi-linear dependence on the $\alpha$-dynamo number $C_{\xi}$ for both poloidal and toroidal components of the magnetic field, with a slope of about $\qty(7\pm 1)\times 10^{-2}$ ms$^{-1}$. Moreover, the dynamo action occurs in a time interval that is much smaller than that on which the evolution of the neutron-finger instability occurs, in support of the hypothesis that the tachocline does not evolve during the action of the dynamo.

To show that the dynamo occurs independently of the initial configuration of the magnetic field and to determine a possible dependence of growth rates on the latter, we perform a further simulation with the lowest value of the dynamo parameter using a purely dipolar (instead than a purely toroidal) initial configuration for the magnetic field, implemented using the solution of Mastrano et al. \cite{Mastrano(2013),Mastrano(2015)}. We find that the initial and final instants of the exponential growth phase are delayed compared to those with the purely tororidal initial configuration. Furthermore, the growth rates are lower than the previous ones.

In the near future we plan to repeat the study for different configurations of the star (lower central rotation rate and/or different equation of state), and/or different profile of the dynamo parameter, and to study the relative importance of the $\alpha\Omega$- and $\alpha^2$-dynamo. We also plan to extend the study to the non-kinematic case, thus taking into account the feedback of the fluid. The study performed in this work can also be extended to hypothetical compact objects such as hybrid stars (i.e., neutron stars with a quark core) and quark stars (i.e., neutron star whose interior is totally composed by quark matter \cite{Drago(2014),Drago(2015),Pili(2016)}.\\

\noindent
\begin{small}
\textbf{Author Contributions:} The authors contributed equally to this work.\\

\noindent
\textbf{Acknowledgments:} The authors thank two anonymous referees for their help in improving the manuscript.\\

\noindent
\textbf{Conflicts of Interest:} The authors declare no conflict of interest.
\end{small}


\begin{thebibliography}{999}
\bibitem[Ozel(2012)]{Ozel(2012)}
\"{O}zel F.; Psaltis D.; Narayan R.; Santos Villarreal A. ON THE MASS DISTRIBUTION AND BIRTH MASSES OF NEUTRON STARS {\em Astrophys. J.} {\bf 2012}, {\em 757}, 55.
\bibitem[Rezzolla(2018)]{Rezzolla(2018)}
Rezzolla L.; Most E. R.; Weih L. R. Using Gravitational-wave Observations and Quasi-universal Relations to Constrain the Maximum Mass of Neutron Stars {\em Astrophys. J.} {\bf 2018}, {\em 852}, L25.
\bibitem[Cromartie(2019)]{Cromartie(2019)}
Cromartie H. T. et al. Relativistic Shapiro delay measurements of an extremely massive millisecond pulsar {\em Nature Astronomy} {\bf 2019}, {\em 4}, 72--76.
\bibitem[Kiuchi(2008)]{Kiuchi(2008)}
Kiuchi L.;Yoshida S. Relativistic stars with purely toroidal magnetic fields {\em Phys. Rev. D} {\bf 2008}, {\em 78}, 044045.
\bibitem[Ciolfi(2009)]{Ciolfi(2009)}
Ciolfi R.; Ferrari V.; Gualtieri L.; Pons J. A. Relativistic models of magnetars: the twisted torus magnetic field configuration {\em Mon. Not. R. Astron. Soc.} {\bf 2009}, {\em 397}, 913--924.
\bibitem[Frieben(2012)]{Frieben(2012)}
Frieben J.; Rezzolla L. Equilibrium models of relativistic stars with a toroidal magnetic field {\em Mon. Not. R. Astron. Soc.} {\bf 2012}, {\em 427}, 3406--3426.
\bibitem[Pili(2014)]{Pili(2014)}
Pili A. G.; Bucciantini N.; Del Zanna L. Axisymmetric equilibrium models for magnetized neutron stars in General Relativity under the Conformally Flat Condition {\em Mon. Not. R. Astron. Soc.} {\bf 2014}, {\em 439}, 3541--3563.
\bibitem[Bucciantini(2015)]{Bucciantini(2015)}
Bucciantini N.; Pili A. G.; Del Zanna L. The role of currents distribution in general relativistic equilibria of magnetized neutron stars {\em Mon. Not. R. Astron. Soc.} {\bf 2015}, {\em 447}, 3278--3290.
\bibitem[Dall'Osso(2009)]{DallOsso(2009)}
Dall'Osso S.; Shore S. N.; Stella L. Early evolution of newly born magnetars with a strong toroidal field {\em Mon. Not. R. Astron. Soc.} {\bf 2009}, {\em 398}, 1869--1885.
\bibitem[Abbott(2017)]{Abbott(2017)}
Abbott B. P. et al. GW170817: Observation of Gravitational Waves from a Binary Neutron Star Inspiral {\em Phys. Rev. Lett.} {\bf 2017}, {\em 119}, 1869--1885.
\bibitem[Uso(1992)]{Uso(1992)}
Uso V. V. Millisecond pulsars with extremely strong magnetic fields as a cosmological source of gamma-ray bursts {\em Nature} {\bf 1992}, {\em 357}, 472--474.
\bibitem[Bucciantini(2009)]{Bucciantini(2009)}
Bucciantini N.; Quataert E.; Metzger B. D.; Thompson T. A.; Arons J.; Del Zanna L. Magnetized relativistic jets and long-duration GRBs from magnetar spin-down during core-collapse supernovae {\em Mon. Not. R. Astron. Soc.} {\bf 2009}, {\em 396}, 2038--2050.
\bibitem[Bucciantini(2012)]{Bucciantini(2012)}
Bucciantini N.; Metzger B. D.; Thompson T. A.; Quataert E. Short gamma-ray bursts with extended emission from magnetar birth: jet formation and collimation {\em Mon. Not. R. Astron. Soc.} {\bf 2012}, {\em 419}, 1537--1545.
\bibitem[Moesta(2015)]{Moesta(2015)}
M\"{o}sta P.; Ott C. D.; Radice D.; Roberts L. F.; Schnetter E.; Haas R. A large-scale dynamo and magnetoturbulence in rapidly rotating core-collapse supernovae {\em Nature} {\bf 2015}, {\em 518}, 376--379.
\bibitem[Ciolfi(2019)]{Ciolfi(2019)}
Ciolfi R.; Kastaun W.; Kalinani J. V.; Giacomazzo B. First 100 ms of a long-lived magnetized neutron star formed in a binary neutron star merger {\em Phys. Rev. D} {\bf 2019}, {\em 100}, 023005.
\bibitem[Duncan(1992)]{Duncan(1992)}
Duncan R. C.; Thompson C. Formation of very strongly magnetized neutron stars - Implications for gamma-ray bursts {\em Astrophys. J.} {\bf 1992}, {\em 392}, L9--L13.
\bibitem[Thompson(1993)]{Thompson(1993)}
Thompson C.; Duncan R. C. Neutron Star Dynamos and the Origins of Pulsar Magnetism {\em Astrophys. J.} {\bf 1993}, {\em 408}, 194--217.
\bibitem[Raynaud(2020)]{Raynaud(2020)}
Raynaud R.; Guilet J.; Janka H.-T.; Gastine T. Magnetar formation through a convective dynamo in protoneutron stars {\em Science Advances} {\bf 2020}, {\em 6}.
\bibitem[Akiyama(2003)]{Akiyama(2003)}
Akiyama S.; Wheeler J. C.; Meier D. L.; Lichtenstadt I. THE MAGNETOROTATIONAL INSTABILITY IN CORE-COLLAPSE SUPERNOVA EXPLOSIONS {\em Astrophys. J.} {\bf 2003}, {\em 584}, 954--970.
\bibitem[Obergaulinger(2009)]{Obergaulinger(2009)}
Obergaulinger M.; Cerd\'{a}-Dur\'{a}n P.; M\"{u}ller E.; Aloy M. A. Semi-global simulations of the magneto-rotational instability in core collapse supernovae {\em Astron. Astrophys.} {\bf 2009}, {\em 498}, 241--271.
\bibitem[Reboul-Salze(2020)]{Reboul-Salze(2020)}
Reboul-Salze A.; Guilet J.; Raynaud R.; Bugli M. A global model of the magnetorotational instability in protoneutron stars {\bf 2020}, arXiv:2005.03567v1.
\bibitem[Bonanno(2003)]{Bonanno(2003)}
Bonanno A.; Rezzolla L.; Urpin V. Mean-field dynamo action in protoneutron stars {\em Astron. Astrophys.} {\bf 2003}, {\em 410}, L33--L36.
\bibitem[Brandenburg(2005)]{Brandenburg(2005)}
Brandenburg A., Subramanian K. Astrophysical magnetic fields and nonlinear dynamo {\em Phys. Rep.} {\bf 2005}, {\em 417}, 1--209.
\bibitem[Naso(2008)]{Naso(2008)}
Naso L.; Rezzolla L.; Bonanno A.; Paternò L. Magnetic field amplification in proto-neutron stars - The role of the neutron-finger instability for dynamo excitation {\em Astron. Astrophys.} {\bf 2008}, {\em 479}, 167--176.
\bibitem[Bucciantini(2013)]{Bucciantini(2013)}
Bucciantini N.; Del Zanna L. A fully covariant mean-field dynamo
closure for numerical 3+1 resistive GRMHD {\em Mon. Not. R. Astron. Soc.} {\bf 2013}, {\em 428}, 71--85.
\bibitem[Parker(1955)]{Parker(1955)}
Parker E. N. Hydromagnetic Dynamo Models {\em Astrophys. J.} {\bf 1955}, {\em 122}, 293.
\bibitem[Bugli(2014)]{Bugli(2014)}
Bugli M.; Del Zanna L.; Bucciantini N. Dynamo action in thick discs around Kerr black holes: high-order resistive GRMHD simulations {\em Mon. Not. R. Astron. Soc.} {\bf 2014}, {\em 440}, L41--L45.
\bibitem[Tomei(2020)]{Tomei(2020)}
Tomei N.; Del Zanna L.; Bugli M.; Bucciantini N. General relativistic magnetohydrodynamic dynamo in thick accretion discs: fully non-linear simulations {\em Mon. Not. R. Astron. Soc.} {\bf 2020}, {\em 491}, 2346--2359.
\bibitem[Del Zanna(2018)]{DelZanna(2018)}
Del Zanna L.; Bucciantini N. Covariant and 3+1 equations for dynamo-chiral general relativistic magnetohydrodynamics {\em Mon. Not. R. Astron. Soc.} {\bf 2018}, {\em 479}, 657--666.
\bibitem[Del Zanna(2016)]{DelZanna(2016)}
Del Zanna L.; Papini E.; Landi S.; Bugli M.; Bucciantini N. Fast reconnection in relativistic plasmas: the magnetohydrodynamics tearing instability revisited {\em Mon. Not. R. Astron. Soc.} {\bf 2016}, {\em 460}, 3753--3765.
\bibitem[Bucciantini(2011)]{Bucciantini(2011)}
Bucciantini N.; Del Zanna L. GRMHD in axisymmetric dynamical
spacetimes - the X-ECHO code {\em Astron. Astrophys.} {\bf 2011}, {\em 528}, A101.
\bibitem[Pili(2017)]{Pili(2017)}
Pili A. G.; Bucciantini N.; Del Zanna L. General relativistic models for rotating magnetized neutron stars in conformally flat space-time {\em Mon. Not. R. Astron. Soc.} {\bf 2017}, {\em 470}, 2469--2493.
\bibitem[Del Zanna(2007)]{DelZanna(2007)}
Del Zanna L.; Zanotti O.; Bucciantini N.; Londrillo P. ECHO: an Eulerian Conservative High Order scheme for general relativistic magnetohydrodynamics and magnetodynamics {\em Astron. Astrophys.} {\bf 2007}, {\em 473}, 11--30.
\bibitem[Miralles(2000)]{Miralles(2000)}
Miralles J. A.; Pons J. A.; Urpin V. A. Convective Instability in Proto–Neutron Stars {\em Astrophys. J.} {\bf 2000}, {\em 543}, 1001--1006.
\bibitem[Mastrano(2013)]{Mastrano(2013)}
Mastrano A.; Lasky P. D.; Melatos A. Neutron star deformation due to multipolar magnetic fields {\em Mon. Not. R. Astron. Soc.} {\bf 2013}, {\em 447}, 3475--3485.
\bibitem[Mastrano(2015)]{Mastrano(2015)}
Mastrano A.; Suvorov A. G.; Melatos A. Neutron star deformation due to poloidal-toroidal magnetic fields of arbitrary multipole order: a new analytic approach {\em Mon. Not. R. Astron. Soc.} {\bf 2015}, {\em 447}, 3475--3485.
\bibitem[Drago(2014)]{Drago(2014)}
Drago A.; Lavagno A.; Pagliara G. Can very compact and very massive neutron stars both exist? {\em Phys. Rev. D} {\bf 2014}, {\em 89}, 043014.
\bibitem[Drago(2015)]{Drago(2015)}
Drago A.; Pagliara G. Combustion of a hadronic star into a quark star: The turbulent and the diffusive regimes {\em Phys. Rev. C} {\bf 2015}, {\em 92}, 045801.
\bibitem[Pili(2016)]{Pili(2016)}
Pili A. G.; Bucciantini N.; Drago A.; Pagliara G.; Del Zanna L. Quark deconfinement in the proto-magnetar model of long gamma-ray bursts {\em Mon. Not. R. Astron. Soc.} {\bf 2016}, {\em 462}, L26--L30.
\end{thebibliography}
\end{document}